%% file: SparCode.tex
\documentclass[sigconf,authorversion]{acmart}

\usepackage{multicol}
\usepackage{multirow}
\usepackage[normalem]{ulem}
\useunder{\uline}{\ul}{}
\usepackage{hyperref}
\usepackage{graphicx}
\usepackage{subfigure}
\usepackage{fixltx2e}
\usepackage{algorithm}
\usepackage{algorithmic}
\usepackage{setspace}
\usepackage{float}
\usepackage{ulem}

\copyrightyear{2023}
\acmYear{2023}
\setcopyright{rightsretained}
\acmConference[SIGIR '23]{Proceedings of the 46th International ACM SIGIR Conference on Research and Development in Information Retrieval}{July 23--27, 2023}{Taipei, Taiwan}
\acmBooktitle{Proceedings of the 46th International ACM SIGIR Conference on Research and Development in Information Retrieval (SIGIR '23), July 23--27, 2023, Taipei, Taiwan}\acmDOI{10.1145/3539618.3591643}
\acmISBN{978-1-4503-9408-6/23/07}

\makeatletter
\gdef\@copyrightpermission{
  \begin{minipage}{0.3\columnwidth}
   \href{https://creativecommons.org/licenses/by/4.0/}{\includegraphics[width=0.90\textwidth]{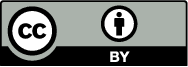}}
  \end{minipage}\hfill
  \begin{minipage}{0.7\columnwidth}
   \href{https://creativecommons.org/licenses/by/4.0/}{This work is licensed under a Creative Commons Attribution International 4.0 License.}
  \end{minipage}
  \vspace{5pt}
}
\makeatother

\begin{document}

\title{Beyond Two-Tower Matching: Learning Sparse Retrievable Cross-Interactions for Recommendation}

\author{Liangcai Su$^*$}
\affiliation{%
  \institution{Shenzhen International Graduate School, Tsinghua University}
  \city{Shenzhen}
  \country{China}
}
\email{sulc21@mails.tsinghua.edu.cn}

\author{Fan Yan}
\affiliation{%
  \institution{Huawei Noah's Ark Lab}
  \city{Shenzhen}
  \country{China}}
\email{yanfan6@huawei.com}

\author{Jieming Zhu$^{\mathsection}$}
\affiliation{%
  \institution{Huawei Noah's Ark Lab}
  \city{Shenzhen}
  \country{China}}
\email{jiemingzhu@ieee.org}

\author{Xi Xiao$^{\mathsection}$}
\affiliation{%
  \institution{Shenzhen International Graduate School, Tsinghua University}
  \city{Shenzhen}
  \country{China}}
\email{xiaox@sz.tsinghua.edu.cn}

\author{Haoyi Duan}
\affiliation{%
  \institution{Zhejiang University}
  \city{Hangzhou}
  \country{China}}
\email{howie@zju.edu.cn}

\author{Zhou Zhao}
\affiliation{%
  \institution{Zhejiang University}
  \city{Hangzhou}
  \country{China}}
\email{zhaozhou@zju.edu.cn}

\author{Zhenhua Dong}
\affiliation{%
  \institution{Huawei Noah's Ark Lab}
  \city{Shenzhen}
  \country{China}}
\email{dongzhenhua@huawei.com	}

\author{Ruiming Tang}
\affiliation{%
  \institution{Huawei Noah's Ark Lab}
  \city{Shenzhen}
  \country{China}}
\email{tangruiming@huawei.com}
\thanks{$^*$ Work done during internship at Huawei Noah's Ark Lab}
\thanks{$^{\mathsection}$ Corresponding authors.}



\renewcommand{\shortauthors}{Liangcai Su et al.}
\begin{abstract}
 Two-tower models are a  prevalent matching framework for recommendation, which have been widely deployed in industrial applications. The success of two-tower matching attributes to its efficiency in retrieval among a large number of items, since the item tower can be precomputed and used for fast Approximate Nearest Neighbor (ANN) search. However, it suffers two main challenges, including limited feature interaction capability and reduced accuracy in online serving. Existing approaches attempt to design novel late interactions instead of dot products, but they still fail to support complex feature interactions or lose retrieval efficiency. To address these challenges, we propose a new matching paradigm named SparCode, which supports not only sophisticated feature interactions but also efficient retrieval. Specifically, SparCode introduces an \textit{all-to-all interaction module} to model fine-grained query-item interactions. Besides, we design a discrete code-based \textit{sparse inverted index} jointly trained with the model to achieve effective and efficient model inference. Extensive experiments have been conducted on open benchmark datasets to demonstrate the superiority of our framework. The results show that SparCode significantly improves the accuracy of candidate item matching while retaining the same level of retrieval efficiency with two-tower models. 
 Our source code will be available at MindSpore/models.

\end{abstract}





\maketitle

\input{sections/1_introduction.tex}
\input{sections/2_realted_work}

\input{sections/3_methods}

\input{sections/4_experiment}

\input{sections/5_conclusion}

\balance
\bibliographystyle{ACM-Reference-Format}
\bibliography{SparCode.bib}


\end{document}

%% file: sections/1_introduction.tex
\begin{figure}[]
	\centering
	\includegraphics[width=0.42\textwidth]{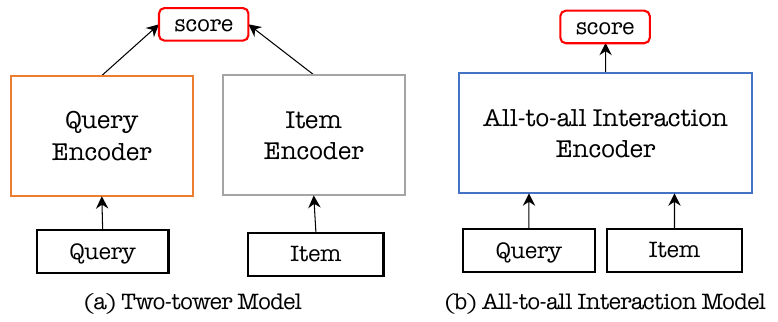}
	\caption{Two tower model and All-to-all Interaction Model.}
	\vspace{-5ex}
	\label{fig_intro}
\end{figure}

\section{Introduction}

{Industrial recommender systems generally consists of candidate matching and ranking phases, where the candidate matching phase plays the role of efficiently retrieving user-preferred items out of a large pool of candidates. The matching result will directly affect the input of the ranking phase, which demonstrates significant importance to the quality of recommender systems. Due to the large size of candidate items (e.g., millions or more), matching models not only need to obtain high accuracy in recall, but also require efficient retrieval to achieve low latency.}

{Two-tower models~\cite{DSSM,MixNS,DAT,SimpleX} are a primary paradigm for candidate matching due to their good accuracy and support for efficient top-\textit{k} retrieval.}
As shown in Figure~\ref{fig_intro}(a), a two-tower model uses dual-encoders (i.e. query and item encoders) to obtain the query\footnote{For recommender systems, user profile is used as a query.} and item representations separately and obtains the final score by simple dot product or cosine computations. 
{For inference, item embeddings can be pre-computed and cached, while only the user embedding is computed online. By utilizing the cached item embeddings and the support of fast Approximate Nearest Neighbor (ANN) search (e.g. Faiss~\cite{Faiss}), two-tower matching has proven to be a good industrial practice in real-world applications~\cite{EBS,DAT,MixNS}}
However, the development of two tower models has the following limitations: \textbf{(a) Limitation 1: Limited feature interaction capability.} There is only one dot product interaction between the two towers, resulting in a limited ability to model fine-grained feature interactions between queries and items. Although dot product is a nice feature interaction method~\cite{NCF_vs_MF}, it is not always optimal, especially in the case when rich content features are available. \textbf{(b) Limitation 2: Reduced accuracy in online serving.} In many industrial scenarios, the size of candidate item pool is so large that exhaustively comparing against all the candidate items becomes impractical due to inference time constraints. ANN search methods (e.g., HNSW~\cite{HNSW}) have been used to build indexes and speed up retrieval, but unfortunately this may reduce recall rate since the model and the index are usually not trained end-to-end.

{To address the above limitations, we propose a new matching paradigm, i.e., SparCode, for improving recommendation accuracy and retrieval efficiency. SparCode consists of two key designs: \textit{all-to-all interaction} and \textit{sparse inverted index}. 
To enhance feature interaction capability, \textit{the all-to-all interaction module} utilizes a single expressive encoder to capture fine-grained interactions between all query features and all item features, as shown in Figure~\ref{fig_intro}(b). It allows different forms of encoder structures (e.g. Self-Attention~\cite{SelfAttention}, MLPs, CrossNet~\cite{DCNv2}), and can learn more fine-grained and predictive feature interactions between the query and the item (i.e., early fusion) than the simple dot product in two-tower models (i.e., late fusion). Fine-grained feature interactions improve model capacity and thus result in improved accuracy, as reported in previous research~\cite{Sentence_Bert,Trans_Encoder}. However, compared with the two-tower models, the nature of query-item encoding makes it hard to use the existing ANN approaches for efficient retrieval, making it impractical. To speed up model inference, a straightforward solution is to pre-compute the scores of user-item pairs and cache them. Thus, SparCode introduces the \textit{sparse inverted index} that is widely used in sparse retrieval. However, there are two problems of building index for users: (1) The number of users is too large to be exhaustive, resulting in an unacceptably expensive precomputation. (2) Each user (index) corresponds to a large number of items, which brings great storage pressure. In respond to these problems, SparCode leverages vector quantization(VQ) as a bridge between all-to-all interaction and sparse inverted indexing, i.e. query is quantized into a series of discrete codes and corresponding code representations, where code replaces query as index and code representation is used for all-to-all interaction. Since the number of codes is controllable and much smaller than the number of queries, the problem of large number of indexes is alleviated. Further, SparCode designs a controlled sparse score function that allows each index to save only the most relevant candidates, greatly relieving storage pressure.}

In summary, we present the advantages of SparCode from both model structure and inference perspectives as follows. {For model structure, SparCode supports sophisticated forms of all-to-all interactions and efficient top-\textit{k} retrieval of large-scale candidates by designing the linkage of VQ and sparse inverted index (for Limitation 1). SparCode introduces all-to-all interaction directly into the model structure, significantly enhancing the ability of feature interaction, rather than using all-to-all interaction models indirectly (e.g. Knowledge Distillation (KD)~\cite{Distilled}) or using parameter-less interaction methods(e.g. ColBERT~\cite{ColBERT}). In addition, sparse inverted indexes are also supported by well-established search tools, such as ElasticSearch, ensuring usability in industrial applications.}

{For model inference, SparCode reduces the gap between model training and inference, since the model and index structure are trained end-to-end (for Limitation 2). In the index structure of SparCode, the index is code, which corresponds to the score of code and candidate items.  The mapping between the query and code, as well as the scores, are learned during model training instead of post-training independently in ANN libraries such as Faiss. Thus, we reduce the performance drop in this stage, which is also empirically verified in our experiments (Sec.~\ref{sec:sparsity_performance_speed}).}

{All in all, SparCode as an all-to-all interaction-based retrieval paradigm, achieves the goal of improving accuracy with the help of enhanced feature interactions while keeping inference efficient. Although we mainly focus on recommender system in this work, SparCode has the potential to be extended to other tasks, such as cross-modal retrieval. We conduct extensive experiments on two public datasets and show that SparCode is significantly more accurate than the two-tower model (Sec. ~\ref{sec:performance_analysis}) as well as comparably efficient to the ANN-based two-tower model (Sec. ~\ref{sec:sparsity_performance_speed}).} Our contributions are summarized as follows.
\begin{itemize}
    \item To the best of our knowledge, SparCode is the first sparse retrieval framework that supports sophisticated forms of all-to-all interactions and controllable sparse inverted indexing for recommender systems.
    \item SparCode converts queries to discrete codes, and thus makes pre-computed scores possible. Besides, 
    SparCode enables efficient retrieval with a sparse inverted index structure, which has mature indexing tool support for industrial deployment. 
    \item Our experiments on public datasets show that SparCode brings a significant improvement in accuracy, while achieving comparable efficiency to the two-tower matching.
\end{itemize}

%% file: sections/2_realted_work.tex
\begin{figure*}[htp!]
	\centering
	\includegraphics[width=0.93\textwidth]{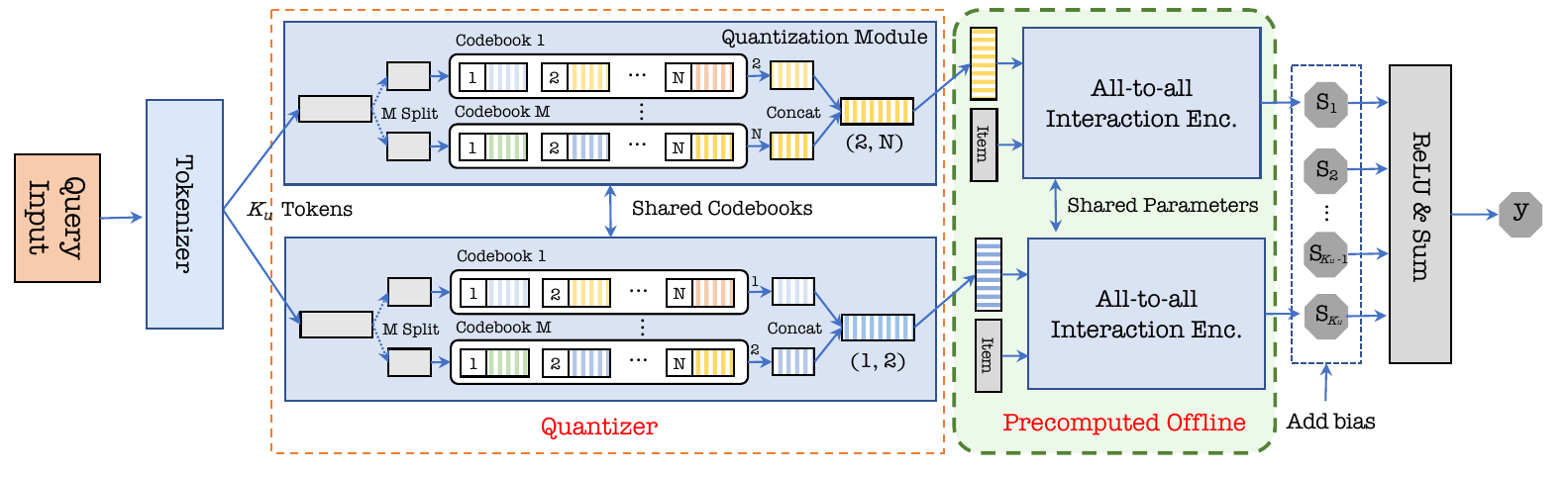}\vspace{-1ex}
	\caption{Overview of SparCode.}
	\vspace{-3ex}
	\label{overview}
\end{figure*}

\section{Related Work}

\subsection{Two-tower Models and Variants}

Two tower matching, the combination of the two-tower model and ANN framework, is the dominant paradigm in dense retrieval, owing to achieving high accuracy and efficient top-\textit{k} retrieval. Two tower matching are widely deployed in various applications~\cite{DSSM,YoutubeDNN,samplingbias, EBS}. 
For these aforementioned partial two-tower models, BARS\cite{zhu2022bars} provides experimental results and leaderboards on multiple datasets for further understanding.
The feature interaction capability between query and item towers is limited by the dot product especially for scenarios with rich features. This has been verified in cross-modal retrieval tasks. For example, in text-image retrieval model VILT~\cite{kim2021vilt}, the all-to-all interaction models performs significantly better than the standard two tower models.  To sum up, enhancing feature interaction between two towers is an important direction for performance improvement. 

There are two variants to overcome the above limitations: (a) using more advanced shallow interactions such as MaxSim~\cite{ColBERT}, attention mechanism~\cite{PolyEncoders}; (b) knowledge distillation, i.e., transferring knowledge from the better interaction-based model to the two-tower model~\cite{Distilled}. However, these variants still do not support \textit{real} all-to-all feature interaction. In addition, some work~\cite{Poeem,GCD} attempts to mitigate the performance loss during inference, using joint training models and indexing. Unlike these methods, SparCode supports sophisticated forms of all-to-all interaction models with superior performance, and uses sparse inverted index instead of the ANN framwork.
\subsection{Vector Quantization}
Vector quantization refers to the discretization of vectors into codes, while product quantization (PQ) is a variant of the sharp increase in the number of codes using VQ multiple times. Traditional Product Quantization employs post-processing to convert vectors into multiple discrete codes for speeding up nearest neighbor search. In recent years, deep quantization has been applied to generating tasks successfully~\cite{VQVAE,VQVAE2,DALL_E} with amazing results. In retrieval task, VQ is also used for multimodal unified modeling~\cite{Cross_Modal_PQ} or representation reconstruction~\cite{PHPQ}.  In general, the code representation obtained by deep quantization is used as an alternative to the query representation to express similar information. 

Different from these works, we use VQ to generate codes for the inverted indexes. The corresponding code representations of the query are used for the all-to-all interaction with the item representations.

%% file: sections/3_methods.tex
\section{Our Methodology}

\subsection{Overview}
In this section, we present our matching framework SparCode. An overview of SparCode is shown in Figure~\ref{overview}. 
To summarize, Tokenizer and Quantizer are designed to obtain multiple discrete representations of a query, supporting sparse inverted indexes for maintaining efficiency. All-to-all interaction-based scorer enhances the interaction between query and item for improving accuracy.

Briefly, we compare the differences between the two-tower model, the all-to-all interaction model, and our SparCode as follows. 
\begin{eqnarray}
\text{Two-tower model:} && score = E_1(q) \circ E_2(c),\label{eq_two_tower} \\ 
\text{All-to-all interaction model:} && score = E(q,c), \label{eq_all}\\ 
\text{SparCode:} && score = \sum_k E\big(\mathcal{T}_k(q),c\big) \label{eq_sparcode},
\end{eqnarray}
where $q$ and $c$ represent a query and item respectively; $E_1(\cdot)$ and $E_2(\cdot)$ are query and item encoder respectively, and $E(\cdot)$ refers to the all-to-all interaction encoder and obtains the score, and $\mathcal{T}(\cdot)$ refers to the Tokenizer and Quantizer, which converts $q$ into a token embedding and looks for codes and alternative code representations.

With these equations, we highlight the design motivation of SparCode. The two tower models (i.e. Eq.~\ref{eq_two_tower}) adopting dot product scoring is to support efficient embedding-based retrieval. The all-to-all interaction models (i.e. Eq.~\ref{eq_all}) supports arbitrarily advanced encoder for fine-grained feature interaction. Usually Eq.~\ref{eq_all} is used for ranking in a few candidates due to its inefficient inference.
SparCode (i.e. Eq.~\ref{eq_sparcode}), based on Eq.~\ref{eq_all}, defines the overall score function as the sum of the scores between each code and candidate. That means that arbitrary queries can be replaced by codes, sharing a code's vocabulary of manageable size. Considering codes as words, the inverted indexes widely used in text retrieval~\cite{BM25} can be easily transferred to the recommendation task. With this definition, SparCode not only introduces a powerful all-to-all interaction but also achieves effecient retrieval similar to sparse retrieval.


\subsection{Matching Framework: SparCode}~\label{Sparcode}


\subsubsection{Tokenizer}
In order to be able to support sparse inverted index, inspired by sparse retrieval, we first encode the query $q$ into multiple token embeddings. The role of the Tokenizer is to represent the query as $K_u$ tokens. Specifically, a query in recommendation (i.e., a user) is encoded into $K_u$ tokens, representing the multiple interests of the user. 

A query may consist of images, text, sequantial information or category features. Uniformly, we represent a query $q\in\mathcal{Q}$ and a candidate item $c\in{\mathcal{I}}$ as a series of embeddings, i.e., 
\begin{equation}
    H^{u}=[h^{u}_{1},h^{u}_{2},...,h^{u}_{L}], H^{u}\in{\mathbb{R}^{L\times{D}}},
\end{equation}
\begin{equation}
    H^{v}=[h^{v}_{1},h^{v}_{2},...,h^{v}_{P}], H    ^{v}\in{\mathbb{R}^{P\times{D}}},
\end{equation}
where $H^{u}$ and $H^{v}$ are the $L$ and $P$ embeddings of a given $q$ and $c$, respectively; $D$ is the dim of each embedding in $H^{u}$ and $H^{v}$. 

Then, we formalize the $i$-th token representation as follows: 
\begin{equation}
    T^{u}_i = Tokenizer(H^{u})\in{\mathbb{R}^{D^T}},
\end{equation}
where $D^{T}$ is the embedding dim of a token depending on the tokenizer settings, usually equals to $D$. 
The specific form of the Tokenizer depends on the given task and features, e.g., if the query is a sequence of historical user clicks, the Tokenizer can be chosen from GRU~\cite{GRU4Rec}, Self-Attention~\cite{SASRec,MIND}, Capsule Network~\cite{MIND}, etc. Similarly, we denote an item as $K_c$ tokens $T^{c}\in{\mathbb{R}^{K_c\times{D^{T}}}}$. 

\subsubsection{Quantizer} 
For text matching tasks, even though the number of queries is very large, they share the same token table (i.e. vocabulary), which limits the number of indexes in the inverted index to the size of the token table.
However, query token embeddings are dense and not shared between queries, making it impossible to build a reasonable number of indexes. Thus, in this section, Quantizer transforms query token embeddings into shared codes and their representations by discretization.

In vector quantization~\cite{VQVAE}, a codebook refers to a series of numbered vectors whose numbers are called codes. The quantification is to input query and return code and the corresponding vector by looking up the codebook, resulting in an arbitrary query sharing the codebook. By utilizing VQ, we quantize token embeddings into discrete codes. We construct $M$ codebooks $C\in{\mathbb{R}^{N\times{\frac{D^{T}}{M}}}}$, each of size $N$, i.e., holding $N$ ordered embeddings. For $T^{u}_i$, we split $T^{u}_i$ into $M$ sub-embeddings $T_{i}^{u,(m)}$ and update it by:

\begin{equation}
\begin{aligned}
    \widetilde{T}^{u,(m)}_{i} & = Quantizer(T_{i}^{u,(m)},C^{(m)}) \\
    & = C^{(m)}_{k}, \text{where} \ k  = \arg\min_{j} \Vert C^{(m)}_{j}-T_{i}^{u,(m)}\Vert_{2},
\end{aligned}
\end{equation}
where $C^{(m)}$ is the $m$-th codebook and $C^{m}_k$ is its $k$-th embedding; $Quantizer$ stands for looking up the most similar sub-token embedding from the given codebook, and the definition of similarity depends on the Euclidean distance between the two sub-embeddings. Thus, we get the complete updated token embedding:
\begin{equation}
    \widetilde{T}^{u}_{i} = \text{Concat}(\widetilde{T}^{u,(1)}_{i}, \widetilde{T}^{u,(2)}_{i},\cdots, \widetilde{T}^{u,(M)}_{i}).
\end{equation}

We use the indexes of $M$ replaced sub-embeddings to combine a corresponding discrete code. For example, in Figure \ref{overview}, suppose that $M=2$ and that the $2$-nd and $N$-th sub-embeddings are taken from $C^{(1)}$ and $C^{(2)}$ respectively, then the discrete code is $(2,N)$. 

There are two considerations for the design of codebook. Firstly, the size $N$ of each codebook should not be set too large since an excessively large codebook affects the speed of our "$Quantizer$". Secondly, $N$ should not be too small since we need enough model capacity to represent different queries.
For the above considerations, we discuss the choice of $M$ and $N$. The number of discrete codes is at most the number $N\times{N}\times{\cdots}\times{N}=N^M$. If $M$ takes 1, no matter how many queries there are, there will ultimately be only $N$ different query embeddings. In order to adequately represent the different queries, $N$ tends to be large, but this increases the number of parameters rapidly and may reduce the speed of quantization. If $M$ is greater than or equal to 2, it is possible to achieve a sufficient number of queries with a small number of parameters and a fast quantization speed.

Furthermore, as "$\arg\min$" is a non-differential operation, there is a non-negligible optimisation problem here. Specifically, the original token embedding ${T}^{u}_{i}$ is unable to obtain the gradient from the updated token embedding $\widetilde{T}^{u}_{i}$, which ultimately results in the previous parameters (e.g., tokenizer) not being updated. We will describe the corresponding model training solution in Section~\ref{optimizer}.


\begin{figure*}[!t]
	\centering
	\includegraphics[width=0.97\textwidth]{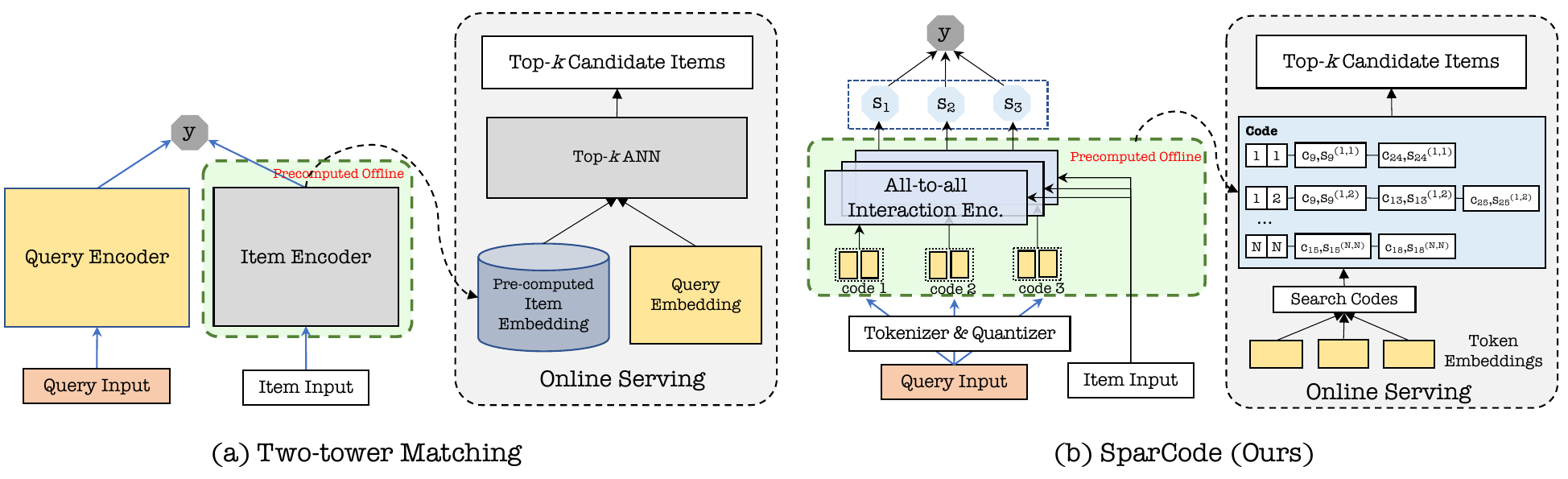}\vspace{-1ex}
	\caption{Comparsion of SparCode and Two-tower Matching.}
	\vspace{-2ex}
	\label{indexing_and_comparsion}
\end{figure*}
\subsubsection{All-to-all Interaction-based Scorer}
As a parameter-free operation, the dot product of the two-tower models accomplishes both feature interaction and scoring between queries and items, which potentially limits the expressiveness of the model.

Different from the previous variants of the two-tower model, we design a parameterised, learnable scorer that supports complex interactions between queries and items, named All-to-all Interaction-based Scorer. 

The proposed scorer does not restrict the specific form of feature interaction, either explicitly (inner product, FM~\cite{FM}, CrossNet~\cite{DCNv2}, Attention~\cite{SelfAttention,PEAR}) or implicitly (e.g. DNN). As an example, We combine inner product and MLPs to give a hybrid scoring function:
\begin{equation}~\label{interaction}
    S_{i} = MLPs([sg[\widetilde{T}^{u}_{i}]\odot{T^{i}_{1}};\cdots;sg[\widetilde{T}^{u}_{i}]\odot{T^{i}_{K_c}}]), i \in{\{1,2,\cdots,K\}}
\end{equation}
where $S_{i}\in{\mathcal{R}^{1}}$ is the matching score between a code and a candidate item $c$; $sg[\cdot]$ represents the stop gradient operation. 
Since codebooks and the rest of the model are optimized separetely, $sg[\cdot]$ is introduced to avoid affecting the parameters of codebooks.
Besides, the query is represented by $K_u$ token embeddings, the above scoring function will get $K_u$ scores separately.

Considering a query or token is only related to a part of candidate items, we
define \textbf{Sparse Score} and \textbf{Final Score} as follow:
\begin{align}
    &\hat{y}_{i} = ReLU(S_{i}+\textbf{b}), \label{sparse_score} \\ 
    &f(q,c)=\hat{y}  = \sum_{i=1}^{K} \hat{y}_{i}, \label{final_score} 
\end{align}
where $\textbf{b}$ is a learnable bias for training.  

The scoring function, especially the part corresponding to Eq.~\ref{sparse_score}, is simple but critical for sparse indexing.
The bias term \textbf{b} of ReLU in Eq.\ref{sparse_score} is the threshold to determine whether to set relevant score $\hat{y}_{i}$ as 0. 
If $\hat{y}_{i}$ equals 0, there is no need to cache $\hat{y}_{i}$ in advance for online serving (See Section~\ref{index_inference} for details).

\subsubsection{Optimization}\label{optimizer}
To make the quantization process training more stable and faster, we use the exponential moving average (EMA) to update codebooks and back-propagation to update the rest of the model like ~\cite{VQVAE2}. In each mini-batch, the parameters in the codebooks and the remainder of the model are updated by the corresponding methods.

\textbf{Model Training.} The model here does not include codebooks. The matching task is the most important objective and SparCode adopts sampled softmax loss for training as follow:
\begin{equation}
    \mathcal{L}_{Match}(q,\mathcal{I}) = \sum_{c\in{\mathcal{I}_{pos}}} log\frac{exp(\hat{y}_{c})}{exp(\hat{y}_{c})+\sum_{\hat{c}\in{\mathcal{I}_{neg}}}exp(\hat{y}_{\hat{c}})},
\end{equation}
where $\mathcal{I}_{pos}$ and $\mathcal{I}_{neg}$ represent positive samples and negative samples sampled for $q$ respectively.


As mentioned earlier, $\arg\min$ is a non-differentiable operation which blocked gradient propagation, causing some parameters such as the tokenizer (or embedding table) cannot get updated.  To update these parameters and make the training more stable, we introduce the commitment loss as follows:
\begin{equation}
    \mathcal{L}_{Commit}(q)=\sum_{i=1}^{K}\sum_{m=1}^{M}\Vert {T}^{u,(m)}_{i}- sg[\widetilde{T}^{u,(m)}_{i}] \Vert_{2}^{2}.
\end{equation}

Thus, the final loss is:
\begin{equation}
    \mathcal{L}(q,\mathcal{I}) = \mathcal{L}_{Match}(q,\mathcal{I}) + \lambda \mathcal{L}_{Commit}(q)
    \label{final_loss},
\end{equation}
where $\lambda$ is a hyperparameter, usually set to 1 or 0.25. 

\textbf{Codebook Update}. 
Following ~\cite{VQVAE2},we update codebooks by EMA, where the embedding in the codebook is is iteratively updated by a combination of itself and the token embeddings mapped to it.  Suppose there is a series of $n^{(s)}_{k}$ sub-embeddings whose nearest sub-embedding is $C^{(m)}_{k}$ in codebook $C^{(m)}$ of the $s$-th mini-batch, we can update $C^{(m)}_{k}$ as:  

\begin{align}
    \mathcal{N}^{(s)}_{k}&:= \mathcal{N}^{(s-1)}_{k} * \gamma + n^{(s)}_{k}*(1-\gamma), \nonumber \\ 
    \mathcal{V}_{k}^{(s)} &:= \mathcal{V}_{k}^{(s-1)}* \gamma+\sum_{j=1}^{n^{(s)}_{k}}T^{u,(m)}(j) * (1-\gamma), \nonumber \\
    C_{k}^{(m),(s)}&:=\frac{\mathcal{V}_{k}^{(s)}}{\mathcal{N}^{(s)}_{k}}, \nonumber
\end{align}
where $\gamma$ is a hyperparameter that adjusts the update rate of the codebook.

\subsection{Indexing and Retrieval}~\label{index_inference}
A query  is transformed into $K_u$ codes by Tokenizer and Quantizer, and each code has a score $s^{code}_c$ with each item $c$ by all-to-all interaction-based scorer. 
Further, the storage cost may be unacceptable if the scores of all ($code, item$) pairs are cached. In most cases, a query is only highly relevant to part of items, which makes sparse inverted index possible since only the score of several top relevant items should be kept for each code. 

Inspired by sparse retrieval and attracted by the efficiency of inverted indexes, we designed a sparse inverted index mechanism for SparCode. We use 0 as a threshold to decide whether to store the score or not for inference.  We rewrite Eq.~\ref{sparse_score} to use a controllable bias instead of the bias term at training time, named "\textbf{Sparsity Control}" as follows: 
\begin{align}
    \hat{y}_{i} & = ReLU(S_{i}+{\widetilde{\textbf{b}}}).
    \label{sparse_control}
\end{align}
Depending on the latency requirements and memory constraints, we adjust $\widetilde{\textbf{b}}$ in Eq.~\ref{sparse_control} to determine the sparsity of indexes.

Figure~\ref{indexing_and_comparsion}(b) illustrates SparCode's online service process. 
The right and blue part shows how the scores are cached, i.e. each code is sparsely cached with the scores of candidate items.It is important to note that these scores are pre-computed.

When serving online, the token embeddings are first obtained by tokenizer, and the code is looked up by codebooks. 
Then, the corresponding item score and filtered candidate itemset are loaded from the cache. For example, in Figure~\ref{indexing_and_comparsion}(a), if the codes are (1,1) and (1,2), we read the score set $\{s_{9}^{(1,1)},s_{24}^{(1,1)},s_{9}^{(1,2)},s_{13}^{(1,2)},s_{25}^{(1,2)}\}$ and get the merged itemset $\{c_9,c_{24},c_{13},c_{25}\}$. Finally, Eq.~\ref{final_score} is used to obtain the final scores and the items with the highest scores are taken as the recommended results. 

\subsection{Comparision with Two-tower Models}~\label{comparision_two_tower}
Figure ~\ref{comparsion_models} shows the comparison between SparCode and Two-tower Matching. We discuss these differences below from two perspectives: modeling stage and inference stage. 

\textbf{Modeling Stage.} 
SparCode supports advanced all-to-all interaction encoders, providing better feature interaction capabilities than two-tower matching, which only supports parameter-free interactions like dot product. Additionally, SparCode's quantizer allows for shared sub-embeddings in different query representations, while each query's representation is independent in the two-tower model.

\textbf{Inference Stage.} As shown in the right part of Figure~\ref{comparision_two_tower}(a) and (b), we  summarize the differences: (1) the cached content. Instead of caching the item embedding, SparCode selectively caches the pre-computed (code, item) scores. (2) the cache structure. SparCode cache is sparse hashing tables for a sparse code-based inverted index, while the two-tower matching caches matrix of item embeddings or other index structures depending on the ANN settings.

%



%% file: sections/4_experiment.tex
\begin{table*}[htbp!]
\setlength{\tabcolsep}{2.5pt}
\centering
\caption{Performance comparisons on Deezer. The second-best results are in bold and the best results are \underline{underlined}.} 
\begin{tabular}{c|ccc|ccc}
\hline
\multirow{2}{*}{\textbf{Methods}}                 & \multicolumn{3}{c|}{\textbf{Deezer (TT-SVD)}}        & \multicolumn{3}{c}{\textbf{Deezer (UT-ALS)}} \\
\multicolumn{1}{l|}{}            & Precision@50 & Recall@50 & \multicolumn{1}{c|}{NDCG@50} & Precision@50     & Recall@50     & NDCG@50     \\ \hline
Popularity                       & 8.92\%       & 3.01\%    & \multicolumn{1}{c|}{9.72\%}  & 8.92\%           & 3.01\%        & 9.72\%      \\
DropoutNet                       & 10.04\%      & 3.75\%    & \multicolumn{1}{c|}{10.46\%} & 16.30\%          & 5.77\%        & 17.62\%     \\
MeLU                             & 15.00\%      & 5.12\%    & \multicolumn{1}{c|}{16.79\%} & 13.92\%          & 4.71\%        & 15.49\%     \\
DeezerNet                          & 9.58\%       & 3.53\%    & \multicolumn{1}{c|}{9.77\%}  & 15.80\%          & 6.63\%        & 20.22\%     \\ \hline
TTM w/ DNN (DSSM)            & 10.40\%      & 3.79\%    & \multicolumn{1}{c|}{10.54\%} & 17.42\%           & 6.09 \%        & 18.60\%      \\
SparCode w/ DNN(ours)         & 17.80\%      & 6.53\%    & \multicolumn{1}{c|}{19.37\%} & 23.45\%           & 8.41\%        & 25.20\%      \\
TTM w/ SA    & 11.22\%      & 4.11\%    & \multicolumn{1}{c|}{11.33\%} & 19.39\%           & 6.93\%        & 21.15\%      \\
SparCode w/ SA(ours) & \textbf{19.08\%}      & \textbf{6.85\%}    & \multicolumn{1}{c|}{ \textbf{20.72\%}} & \textbf{24.13\%}           & \textbf{8.69\%}        & \textbf{25.99\%}      \\ \hline
All-to-All SA & {\ul 19.17\%} & {\ul 6.96\%} & \multicolumn{1}{c|}{{\ul 20.75\%}} & {\ul 26.35\%} & {\ul 9.65\%} & {\ul 28.23\%}   \\\hline
\end{tabular}
\label{performance_deezer}
\end{table*}


\begin{table}[htbp!]
\setlength{\tabcolsep}{2.5pt}
\centering
\caption{Performance comparisons on Movielens-10M. The second-best results are in bold and the best results are \underline{underlined}.} 

\begin{tabular}{c|ccccccccccc}
\hline
Model Type & Model & Precision@50 & NDCG@50  & Recall@50 \\ \hline
\multirow{3}{*}{TTM} & DSSM & 8.124\% & 17.690\% & 26.534\% \\
 & GRU4Rec & 8.356\% & 18.473\% & 28.261\% \\
 & SASRec & 8.754\% & 19.202\% & 29.828\% \\ \hline
\multirow{3}{*}{SparCode} & DNN  & 8.393\% & 18.295\% & 27.184\% \\
 & GRU4Rec & { 8.888\%} & { 20.050\%} & { 29.492\%} \\
 & SASRec  & \textbf{ 9.020\%} & \textbf{ 20.480\%} & \textbf{ 30.142\%} \\ \hline
 All-to-All & SASRec & {\ul 9.378\%} & {\ul 21.950\%} & {\ul 31.946\%} \\\hline
\end{tabular}
\label{performance_ml}
\end{table}

\section{Experiment}

\subsection{Experimental Setup}
\subsubsection{Datasets}
We conduct experiments on two datasets {Deezer}\footnote{\url{https://zenodo.org/record/5121674\#.YwpGvC-KFAa}} and {Movielens-10M}~\footnote{\url{https://grouplens.org/datasets/movielens/}} for candidate item matching. 

\textbf{Deezer} dataset contains 100,000 fully anonymous users and 50,000 music tracks. We follow ~\cite{Deezer}, using 70,000 active warm users for training and the remaining 20,000 and 10,000 cold users for validation and testing. The dataset provides two pretrained embedding types named "TT-SVD" and "UT-ALS" to represent users, user features and songs. 

\textbf{Movielens-10M} dataset is a classic recommendation dataset, consisting of 71,567 users and 65,133 items and over 10,000,000 user interactions. It contains abundant sequential user interactions and is widely evaluated for sequential item retrieval.  
We split all users into training, validation and test sets by the ratio of 8:1:1.

\subsubsection{Competitors}
To evaluate the results, our proposed method is compared with several powerful baselines in recent literature. For simplicity, we abbreviate the "Two-Tower Model" as "TTM" for all the tables and figures. 

For {Deezer}, we adopt a popularity-based method called "Popularity" and three models specialized in cold-start recommendations include DropoutNet~\cite{DropoutNet}, MeLU~\cite{MeLU}, and DeezerNet~\cite{Deezer}. In addition, we compare SparCode with two two-tower models: DSSM~\cite{DSSM} and SA~\cite{SASRec}, in which DSSM is a famous two-tower model based-on DNN while SA utilizes Self-Attention module as encoder for feature encoding. We denote DSSM as ``TTM w/ DNN'' and SA as ``TTM w/ SA''. For SparCode, we choose DNN and Self-Attention modules as its tokenizer, denoted as ``SparCode w/ DNN'' and ``SparCode w/ SA'' respectively. 

For {Movielens-10M}, we choose three most commonly used baselines including DSSM~\cite{DSSM}, GRU4Rec~\cite{GRU4Rec} and SASRec~\cite{SASRec}. The sequential modeling module of these methods are treated as the user tower.  
Respectively, we choose DNN, GRU, and Self-Attention modules as tokenizer for SparCode to evaluate on this dataset. 
We set the length of the behaviour sequence to 20 for all methods.

For both recommendation datasets, we choose vanilla All-to-All Self-Attention model as a strong baseline, denoted as ``All-to-All SA''. Without consideration of inference speed, the user features/histories and item features are fed into a mulit-layer SA-based model simultaneously for better feature interaction. The performance gap between SparCode and All-to-All SA reveals how much effectiveness sacrificed by SparCode for better efficiency. 


\subsubsection{{Implementation Details}}~\label{app_implement_details}
We choose Adam~\cite{Adam} as optimizer with the learning rate of 0.001. We set the $L_2$ regularization factor for the embedding table as $1e^{-6}$ and the dropout rate as 0.1. Uniformly, the hidden units of DNNs is [256,256,256], and Self-Attention layers is 3. For SparCode, we search for the best query token number $K_u$ from $\{1,2,4,6,8,10\}$. The different $K_u$ query token embeddings are obtained from SENet~\cite{SENet} or linear layers based on the output f the Tokenizer. 
 The different $K_c$ item token embeddings are encoded from different linear layers. 
The hyperparameters of the codebooks are of vital importance, including codebook number $M$, codebook capacity $N$, and the size of each embedding. 
We do grid search of $M$ from $\{1,2\}$ and $N$ from $\{64,128,256,512,1024\}$. The item embedding size is set as 128 and 256 for "TT-SVD" and "UT-ALS" respectively on Deezer, and 64 on Movielens-10M. $\lambda$ in Eq.~\ref{final_loss} is choosen from $\{0.25,1\}$. To ensure the accuracy of the results, we repeated the experiment five times for each experiment with different random seeds. Our implementation is partially based on FuxiCTR~\cite{FuxiCTR}, a library for CTR prediction. 

\subsection{Performance Analysis}~\label{sec:performance_analysis}
A comparison of the performance of SparCode and baselines is shown in the Table~\ref{performance_deezer} and ~\ref{performance_ml}. As shown in Table~\ref{performance_deezer} and ~\ref{performance_ml}, SparCode performs well on both datasets, far better than two-tower models and even close to the "oracle", the All-to-All SA model. Specifically, SparCode w/ DNN achieves 71.15\% relative improvements over DSSM for Precision@50 on Deezer. Compared with other state-of-the-art retrieval models such as MeLU, DropoutNet, and DeezerNet, SparCode also yields significant effectiveness gain. For Movielens-10M, SparCode outperforms all tow-tower-models with the same sequential encoder on various metrics. For SparCode, we found that self-attention module is more suitable to be used as the tokenizer to encode interacted feature fields or sequential user behaivors. 

Since SA-based module performs best on both datasets, the All-to-All SA model acts like an ``Oracle'' which shows the performance upper bound of SparCode. Specifically, we remove the PQ and sparse indexing modules in SparCode with SA for both datasets.  
The results are shown as "All-to-All SA" for Deezer and "All-to-All SASRec" for Movielens-10M. For TT-SVD Embedding based Deezer, the performance gap between SparCode with SA and All-to-All SA is only 0.4\% for precision@50. For UT-ALS Embedding based Deezer and Movielens-10M, the performance gap is slightly larger but is also acceptable. For instance, the relative gap on Recall@50 between SparCode w/ SA and All-to-All SA is  11\%. This indicates that the use of PQ and sparse index will not significantly degrade performance. SparCode achieves a nice trade-off between effectiveness and efficiency.

\subsection{Abalation Study}~\label{sec:abalation_study}
\begin{table}[]
\setlength{\tabcolsep}{2.5pt}
\centering
\caption{Effect of Interaction Type on Deezer with "TT-SVD" Embeddings. R: Recall, N: NDCG.}
\resizebox{0.88\linewidth}{!}{
\begin{tabular}{c|c|ccccc}
\hline
Model                               & Interaction Type              & R@20       & R@50       & N@20          & N@50          \\ \hline
TTM w/ DNN                      & {Dot Product}        & 1.58\%          & 3.79\%          & 10.92\%          & 10.54\%          \\ \hline 
\multirow{6}{*}{\shortstack{SparCode\\ w/ DNN}} & Dot Product($K_c$=1)                  & 1.40\%          & 3.39\%          & 9.61\%           & 9.39\%           \\
                                    & Dot Product($K_c$=4)                  & 1.45\%          & 3.51\%          & 9.93\%           & 9.69\%           \\
                                    & MaxSim                              & 2.46\%          & 5.44\%          & 17.22\%          & 15.68\%          \\
                                    & CrossNet                       & 2.42\%          & 5.30\%          & 17.33\%          & 15.60\%          \\
                                    & DNN                             & 3.12\%          & 6.55\%          & 21.45\%          & 18.27\%          \\
\multicolumn{1}{l|}{}               & InnerPDNN                            & {\ul 3.18\%}    & 6.53\%          & {\ul 23.37\%}    & 19.96\%          \\ \hline
TTM w/ SA                       & {Dot Product}            & 1.70\%          & 4.11\%          & 11.63\%          & 11.33\%          \\ \hline
\multirow{6}{*}{\shortstack{SparCode\\ w/ SA}}                   & Dot Product($K_c$=1)                  & 1.55\%          & 3.72\%          & 10.71\%          & 10.34\%          \\
                                    & Dot Product($K_c$=4)                  & 1.57\%          & 3.75\%          & 10.77\%          & 10.43\%          \\
                                    & MaxSim                              & 2.39\%          & 5.42\%          & 16.67\%          & 15.47\%          \\
                                    & CrossNet                      & 2.49\%          & 5.55\%          & 17.55\%          & 16.11\%          \\
                                    & DNN                            & {\ul 3.18\%}    & {\ul 6.63\%}    & 23.20\%          & {\ul 20.09\%}    \\
\multicolumn{1}{l|}{}               & InnerPDNN                           & \textbf{3.32\%} & \textbf{6.85\%} & \textbf{24.11\%} & \textbf{20.72\%} \\ \hline
\end{tabular}%
}
\label{tbl_interaction_type}
\end{table}
\subsubsection{Interaction Type}
We explore the impact of different types of feature interactions on SparCode. 
Considering that we model query and item as multiple token representations, we give a generic form for different feature interactions by rewriting Eq.~\ref{interaction} as follows.
\begin{itemize}
    \item \textbf{Dot Product}: $S_i = \sum_j^{p}
    \langle sg[\widetilde{T}^{u}_{i}],{T^{i}_{K_c}} \rangle$.
    \item \textbf{MaxSim}(ColBERT~\cite{ColBERT}): $S_i = {Max}_{i\in{1,2,\odot,p}}(
    \langle sg[\widetilde{T}^{u}_{i}],{T^{i}_{K_c}} \rangle)$.
    \item \textbf{CrossNet}~\cite{DCNv2}: $S_i = {CrossNet}(
    [ sg[\widetilde{T}^{u}_{i}];{T^{i}_{1}};\cdots;{T^{i}_{K_c}} ])$.
    \item \textbf{DNN}: $S_i = {MLPs}(
    [ sg[\widetilde{T}^{u}_{i}];{T^{i}_{1}};{T^{i}_{2}};\cdots;{T^{i}_{K_c}} ])$.
    \item \textbf{InnerPDNN}: \\ $S_{i} = MLPs([sg[\widetilde{T}^{u}_{i}]\odot{T^{i}_{1}};\cdots;sg[\widetilde{T}^{u}_{i}]\odot{T^{i}_{K_c}}])$ (i.e. Eq.~\ref{interaction}).
    
\end{itemize}

Among these interaction ways, Dot Product and MaxSim (from ColBERT~\cite{ColBERT}) are parameterless methods while the other three have optimizable parameters. The number of CrossNet layers is set as 3. InnerPDNN is the interaction way shown in Eq.~\ref{interaction}, which is the combination of Inner Product and DNN.  Table~\ref{tbl_interaction_type} shows the experimental results of various interaction types on Deezer with "TT-SVD" Embeddings. "TTM w/ DNN" is the same as DSSM. 



According to Table~\ref{tbl_interaction_type}, complex interaction approaches (e.g., DNN, MaxSim) achieve significant performance gains compared with the simple dot production. Specifically, for DNN-based models, SparCode with MaxSim improves from 3.39\% to 5.44\% for Recall@50 and from 9.39\% to 15.68\% for NDCG@50 compared with SparCode with Dot Product. Similar phenomenon can be observed from the SA-based models, with a 45.69\% and 49.61\% relative improvement in Recall@50 and NDCG@50 respectively. The above results indicate that the dot product may not be sufficient to model the interaction between query and item.  The DNN interaction performs significantly better than MaxSim and CrossNet. 
For SparCode with SA, the DNN interaction improves the Recall@50 metric from MaxSim's 5.42\% and CrossNet's 5.55\% to 6.63\%, which is a remarkable relative improvement.  
Since Deezer dataset is rich in content semantics but not restricted to sparse ID features, the cross feature network may not as good as DNN to model such implict query-item interaction. 

Another interesting observation is the performance of SparCode with dot product interaction is always slightly lower than that of two-tower models, suggesting that the replaced vectors from codebooks compromise some performance for efficiency. We leave how to reduce the performance gap as a future research topic.  


\subsubsection{Codebook Structure}

\begin{figure}[t!]
	\centering
	\subfigure[$M$=1 on Deezer] {\includegraphics[width=.23\textwidth]{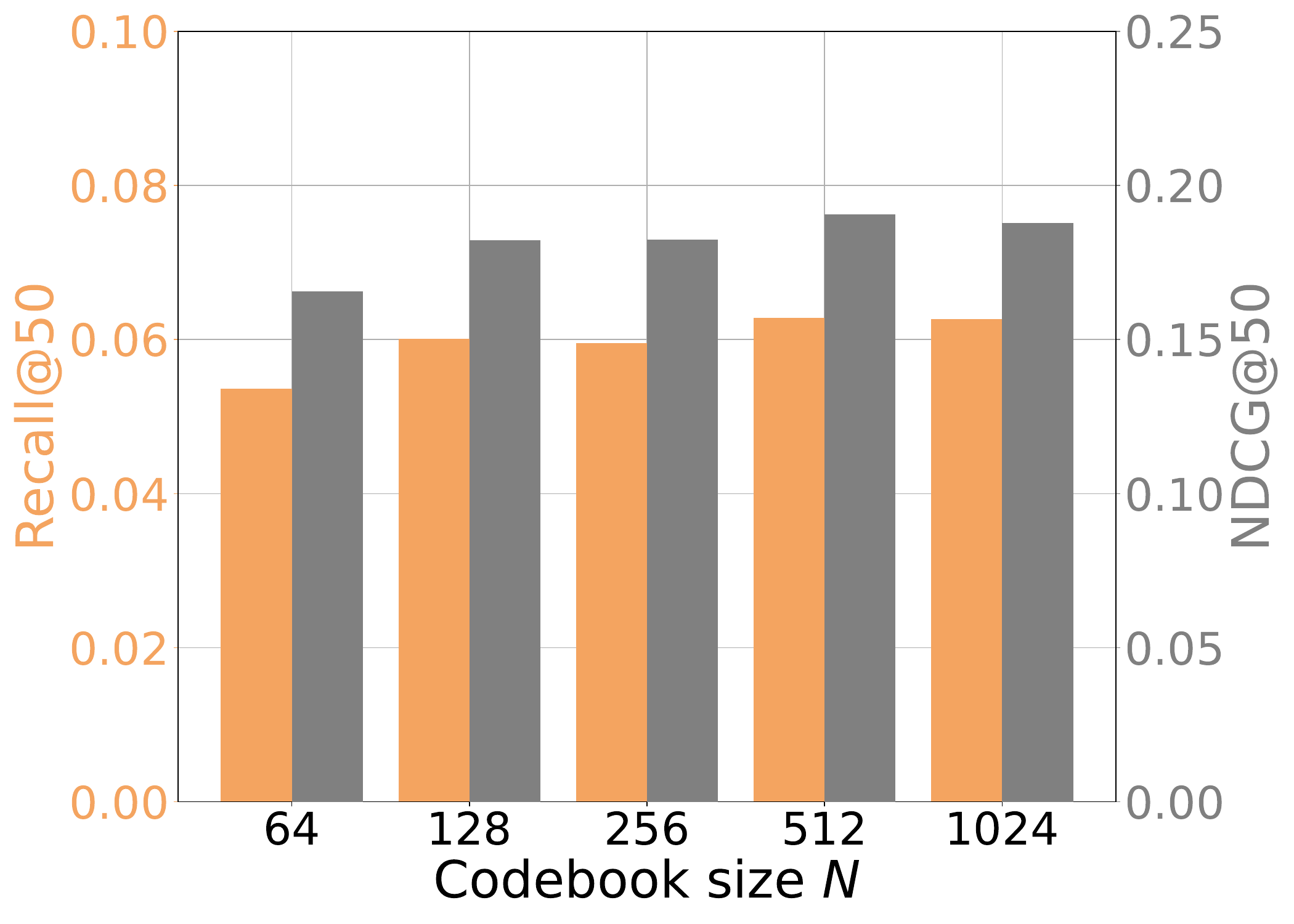}}
	\subfigure[$M$=2 on Deezer] {\includegraphics[width=.23\textwidth]{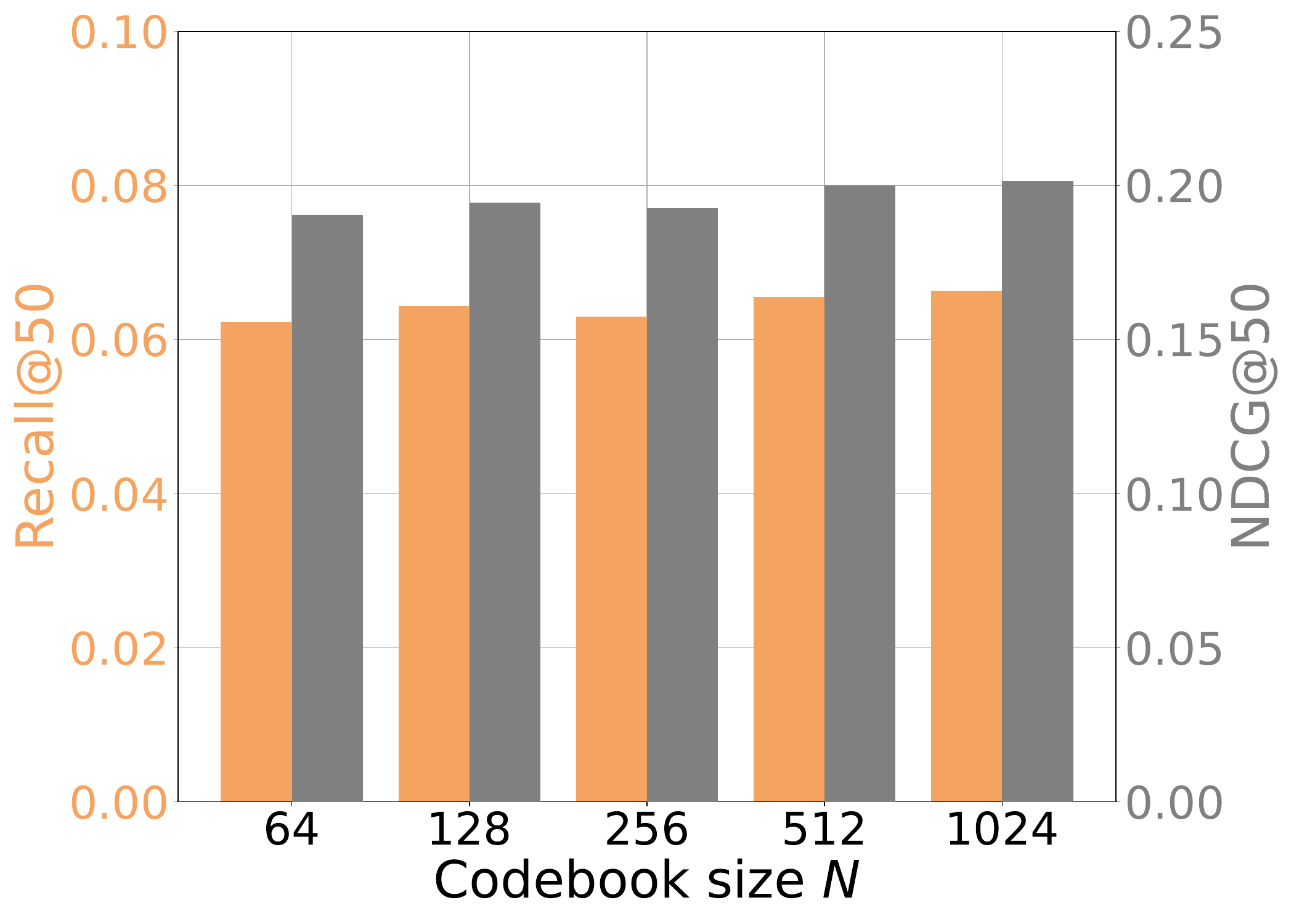}}
	\subfigure[$M$=1 on Movielens-10M] {\includegraphics[width=.23\textwidth]{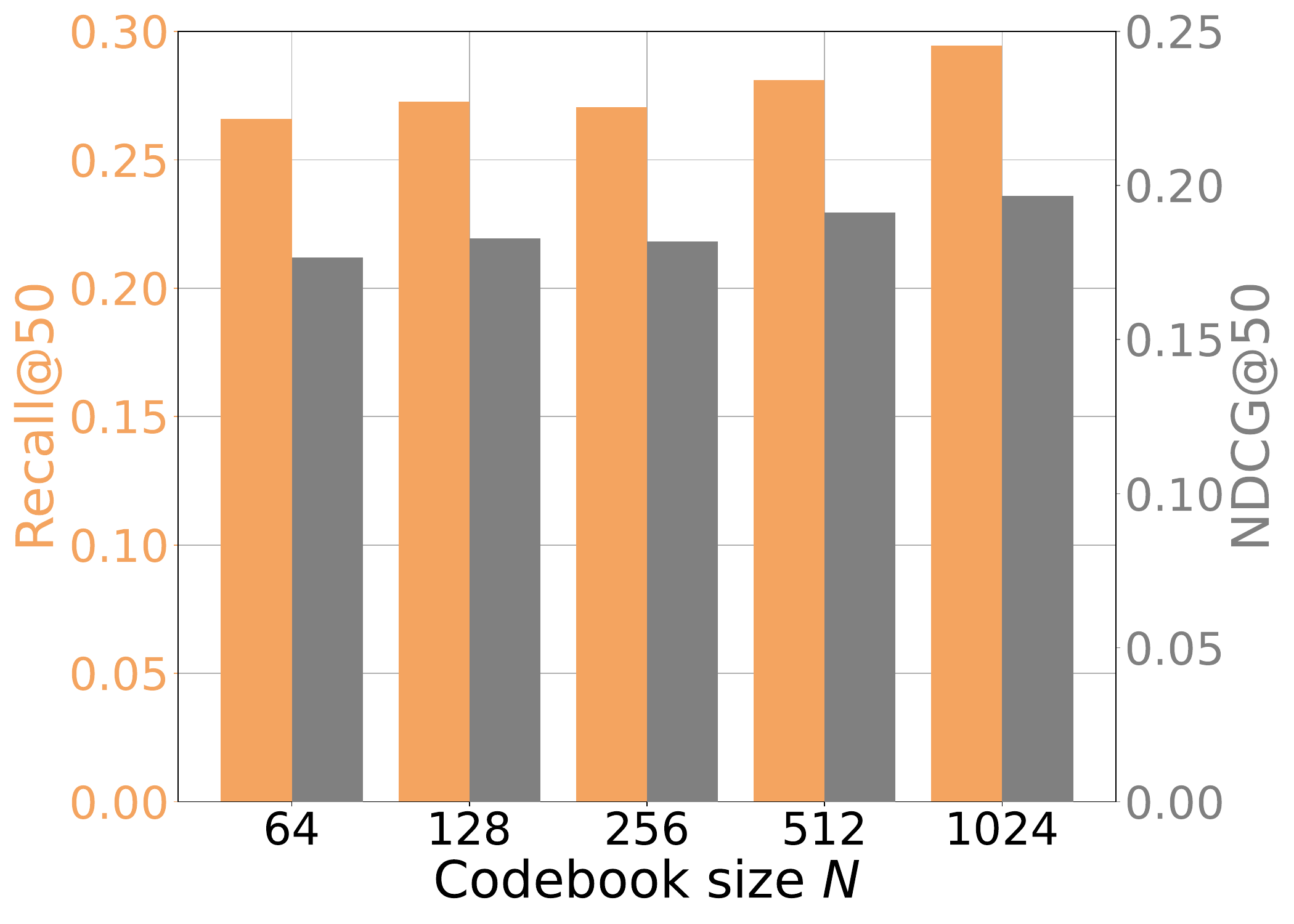}}
	\subfigure[$M$=2 on Movielens-10M] {\includegraphics[width=.23\textwidth]{figures/Movielens-10M_SA_codebookN_2.pdf}}
    \vspace{-2ex}
	\caption{The number and capacity of codebook(s).}
	\label{fig_codebook}
     \vspace{-2ex}
\end{figure}




Tuning codebooks is a key step in PQ. We explored the impact of hyperparameters related to the number and capacity of codebooks. We chose SparCode w/ DNN on Deezer and SparCode w/ SA on Movielens-10M for experiments and select Recall@50 and NDCG@50 as evaluation metrics. According to the results in Figure~\ref{fig_codebook}, we have the following observations.

Firstly, increasing the capacity $N$ of a single codebook is helpful for better performance. On Deezer, as the codebook size increases from 64 to 1024, both Recall and NDCG are on the rise. When there is only one codebook($M$=1), Recall@50 and NDCG@50 get improved by 16.92\% and 13.29\% when $N$ increases from 64 to 1024. When $M$ is 2, the improvement is 6.53\% and 5.84\%. This is because the larger $M$ brings more code words combinations, which enhanced the representation of PQ thus the relative performance boost is not so significant. 
To conclude, it is recommended to increase the capacity of the codebook size for better performance, especially when there is only one codebook.


Secondly, the number of codebooks $M$ has tremendous effects of final performance. As shown in Figure~\ref{fig_codebook}(a)(b), comparing $M$=1, $N$=1024 with $M$=2, $N$=64, the latter one is already more effective than the former. We believe the shared information between codebooks contributes to the increase. When $M$>1, different queries may share the same codewords in a specific codebook. Thus the learned information are shared in this codeword. We believe such representation sharing potentially prevents overfitting. However, it is worth noting that increasing $M$ brings cost on computing. Choosing a proper $M$ is also a trade-off between efficiency and effectiveness.  



\subsubsection{{Effect of $K_u$}}
\begin{figure}[t!]
	\centering
	\subfigure[Deezer]
	{\includegraphics[width=0.24\textwidth]{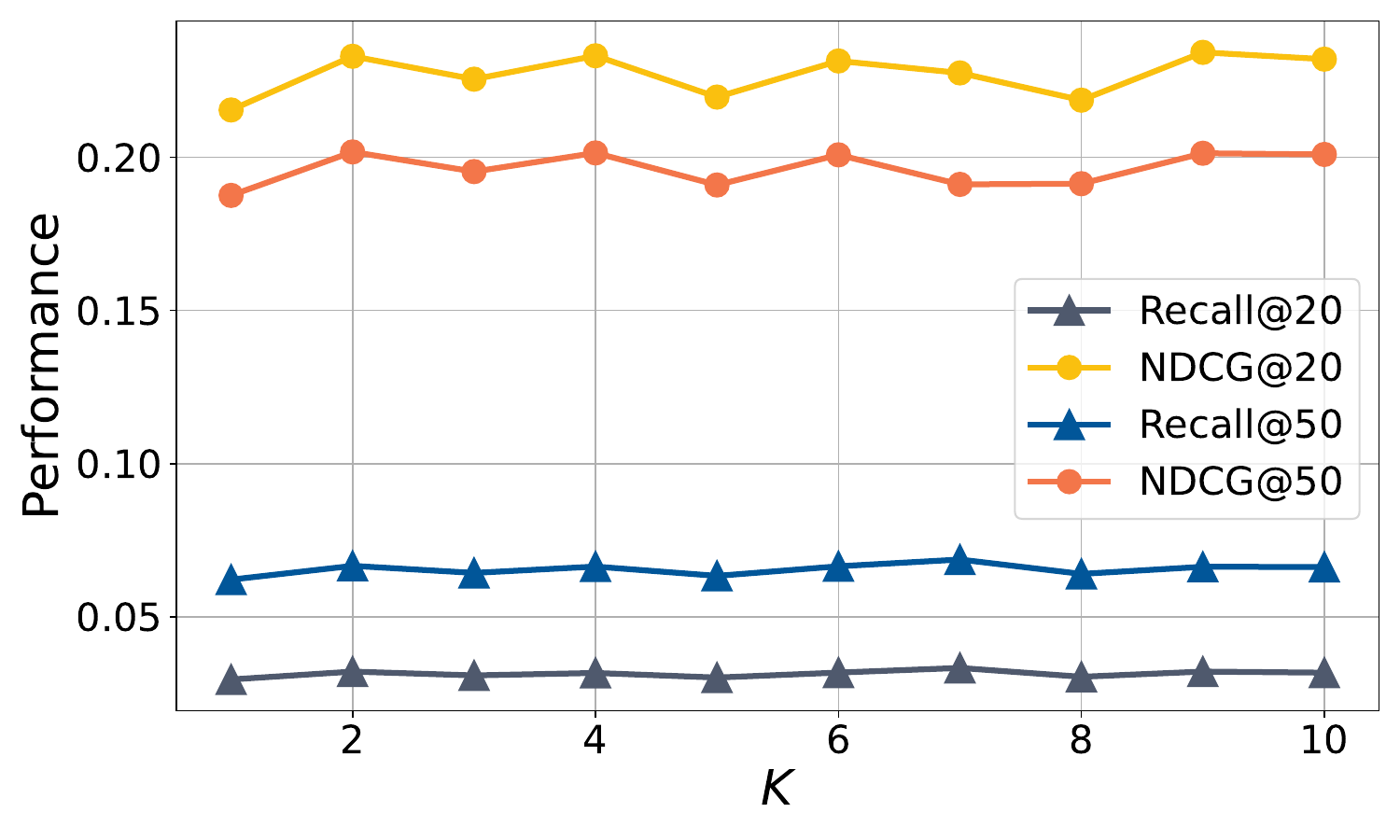}}
	\subfigure[Movielens-10M]
	{\includegraphics[width=0.24\textwidth]{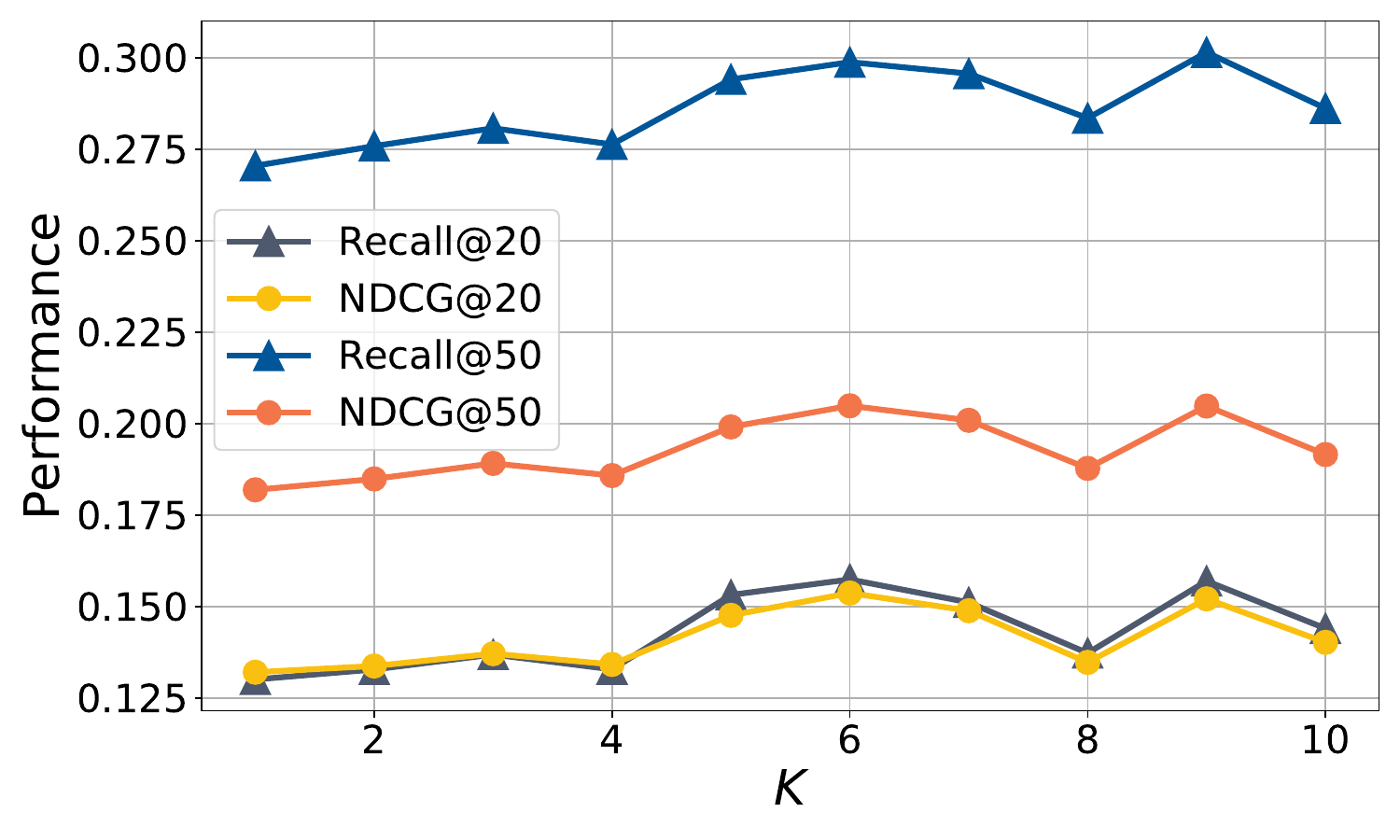}}
    \vspace{-2ex}
	\caption{Effect of $K_u$.}
	\vspace{-4ex}
	\label{comparsion_models}
\end{figure}

The tokenizer of SparCode encode query and item into $K$ token vectors for later interaction. We utilize learnable linear layers or SENet~\cite{fibinet} to transform various inputs into fixed tokens. We investigates the effects of different hyperparameter $K_u$. The results is shown in Fig~\ref{comparsion_models}. For sequential recommendation dataset like MovieLens-10M, the increase of K brings performance on various metrics, which indicates various encoded tokens represented multiple user interests, and is beneficial for better interests matching. For non-sequential recommendation dataset such as Deezer, such improvements is not very clear. Thus we can set a small $K_u$ (e.g. $K_u=2$) for better efficiency.

\subsection{Sparsity, Performance and Speed}~\label{sec:sparsity_performance_speed}
Sparse Indexing Mechanism play a key role in SparCode deployment. In this section, we explore the impact of cache sparsity on model performance and efficiency. 
To be more intuitive, we provide two evaluation metrics for sparsity, "\textit{Sparsity}" and "\textit{Average Items}". Specifically, "\textit{Sparsity}" represents the ratio of \textbf{not}-cached code-item scores to the total scores. "\textit{Average items}" represents the average number of cache scores per code. These two metrics are formulated as follows:
\begin{gather}
    \textit{Sparsity}  = 1 -  \frac{\#(s^{(m,n)}_{c}>0)}{M^N \times |\mathcal{I}|}, \nonumber \\
    \textit{Average Items}  = \frac{\#(s^{(m,n)}_{c}>0)}{M^N}, \nonumber
\end{gather}
where $m\in{\{1,2,\cdots,M\}}, n\in{\{1,2,\cdots,N\}}$ and "\#" means "the number of".

\begin{figure*}[t!]
	\centering
	\subfigure[Effects of sparsity on performance for Deezer]
	{\includegraphics[width=.43\textwidth]{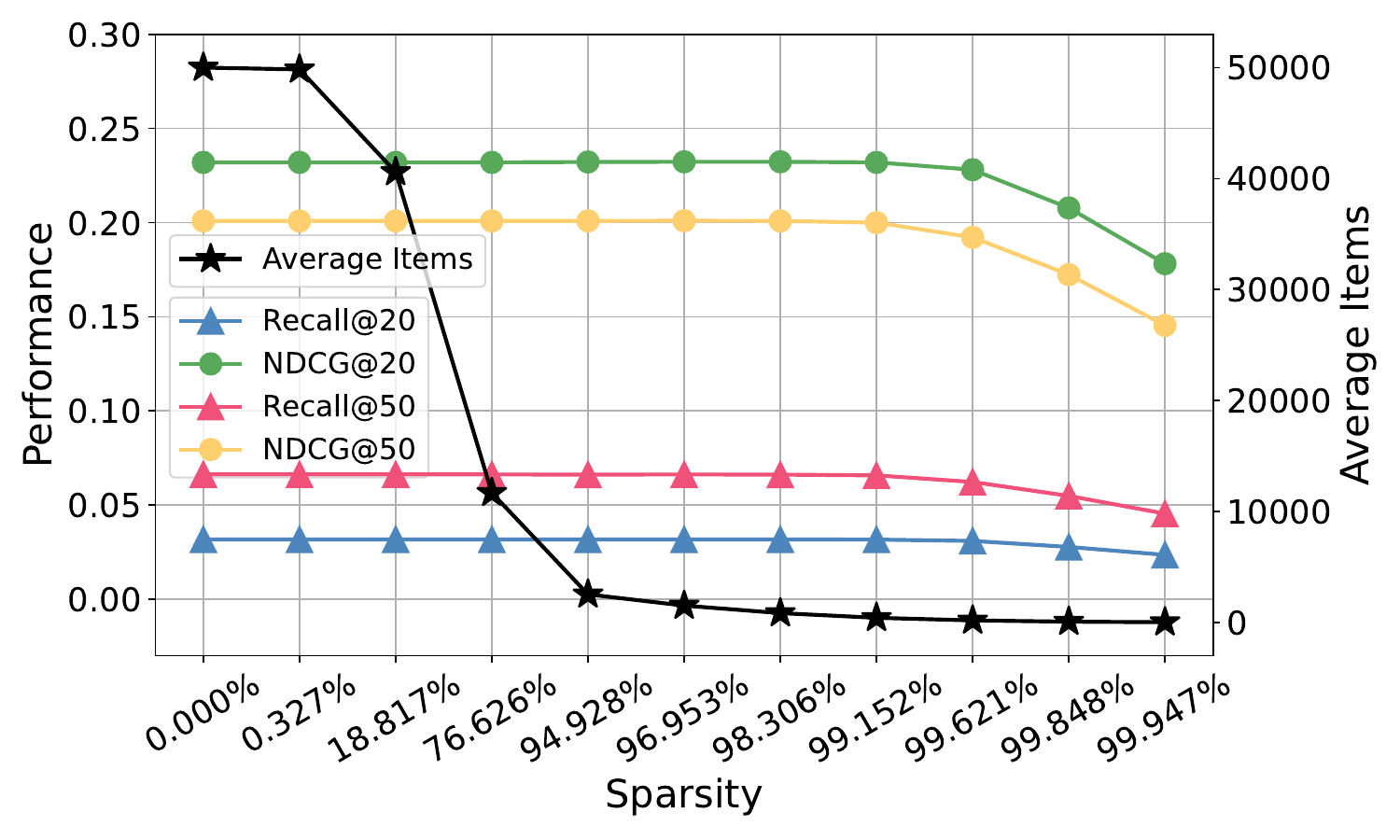}}
	\subfigure[Effects of sparsity on performance for Movielens-10M] {\includegraphics[width=.43\textwidth]{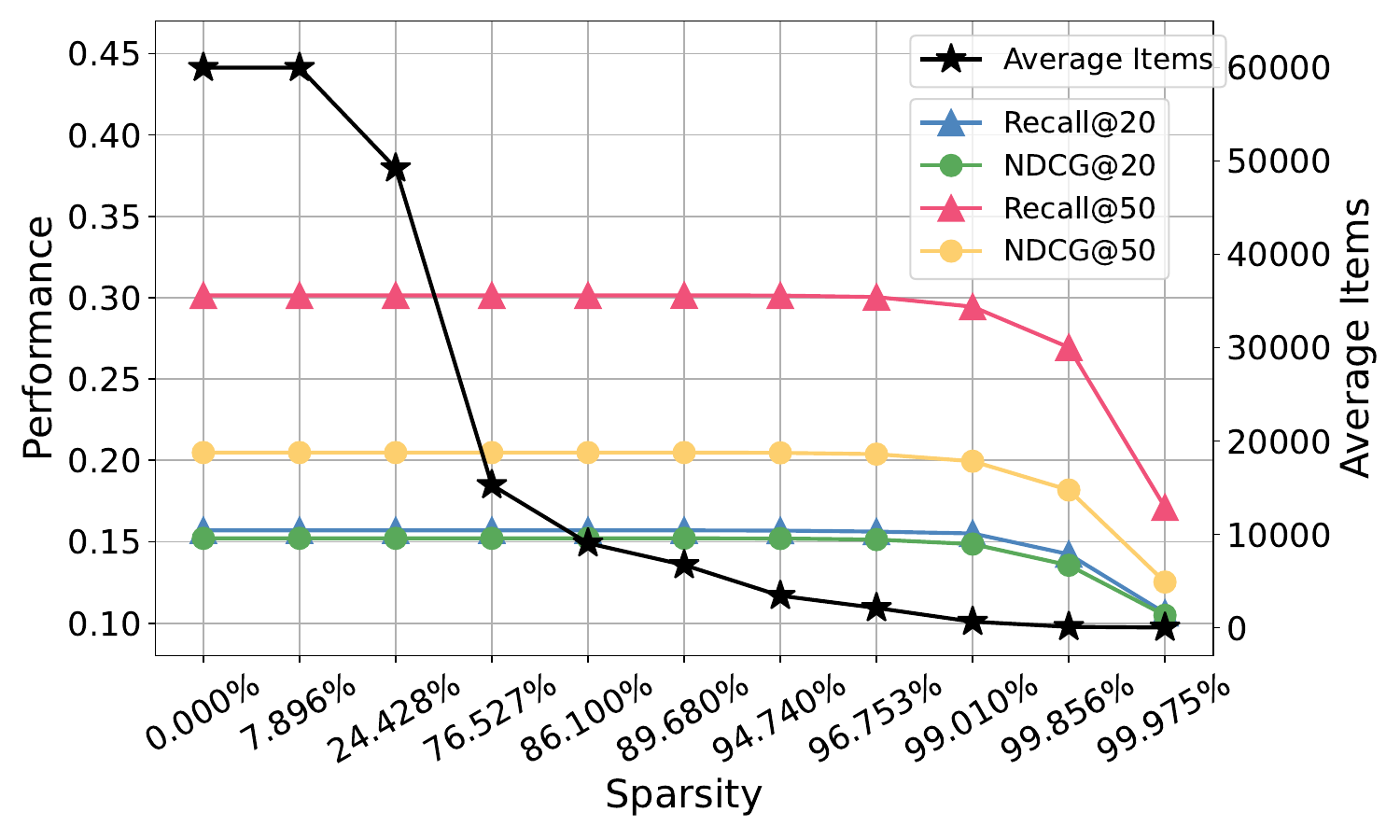}}
	\subfigure[Comparison of query time of SparCode and TTMs. The statics annotation denotes sparsity.] 
	{\includegraphics[width=.42\textwidth]{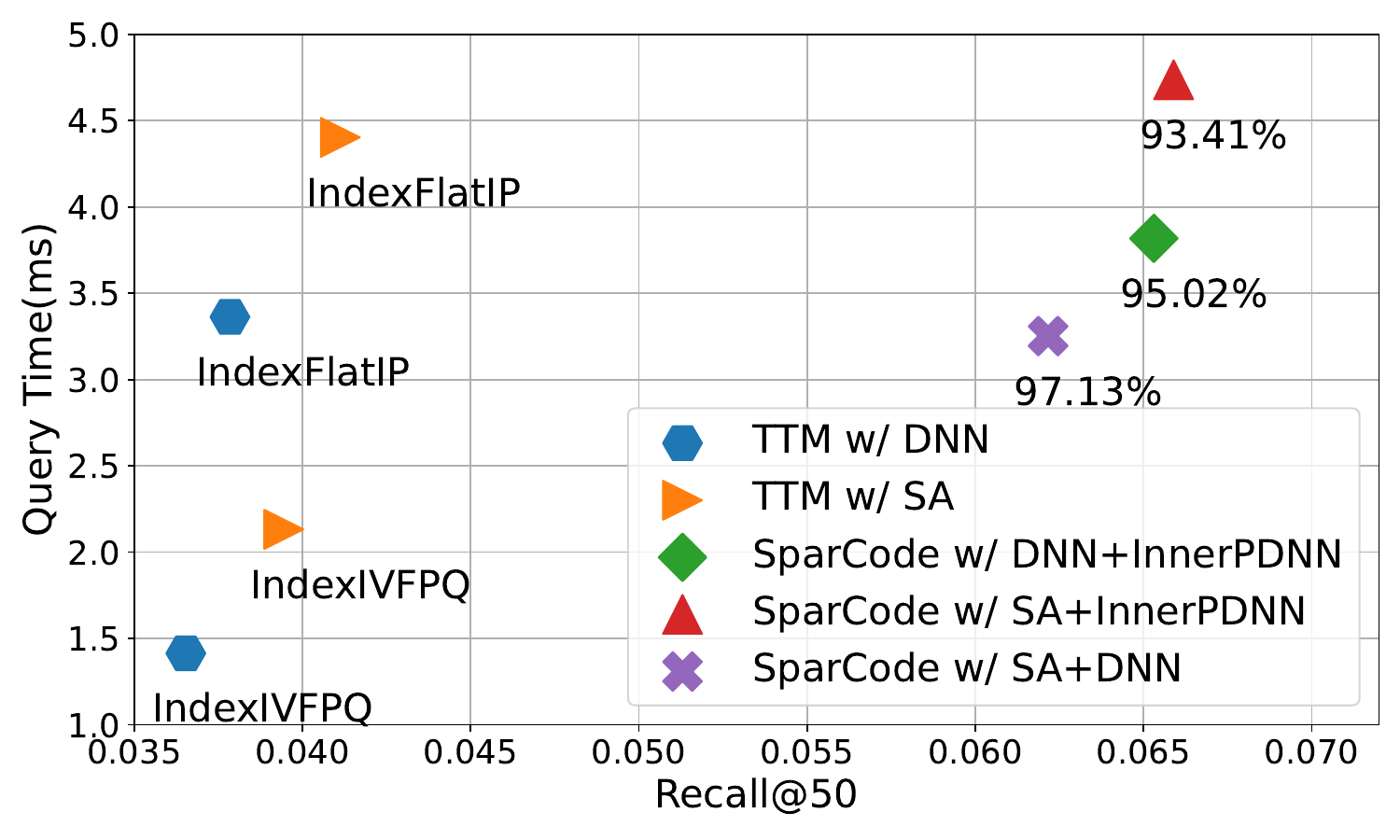}}
	\subfigure[Comparison of query time under different Sparsity. The statics above the points denotes Recall@50.] {\includegraphics[width=.42\textwidth]{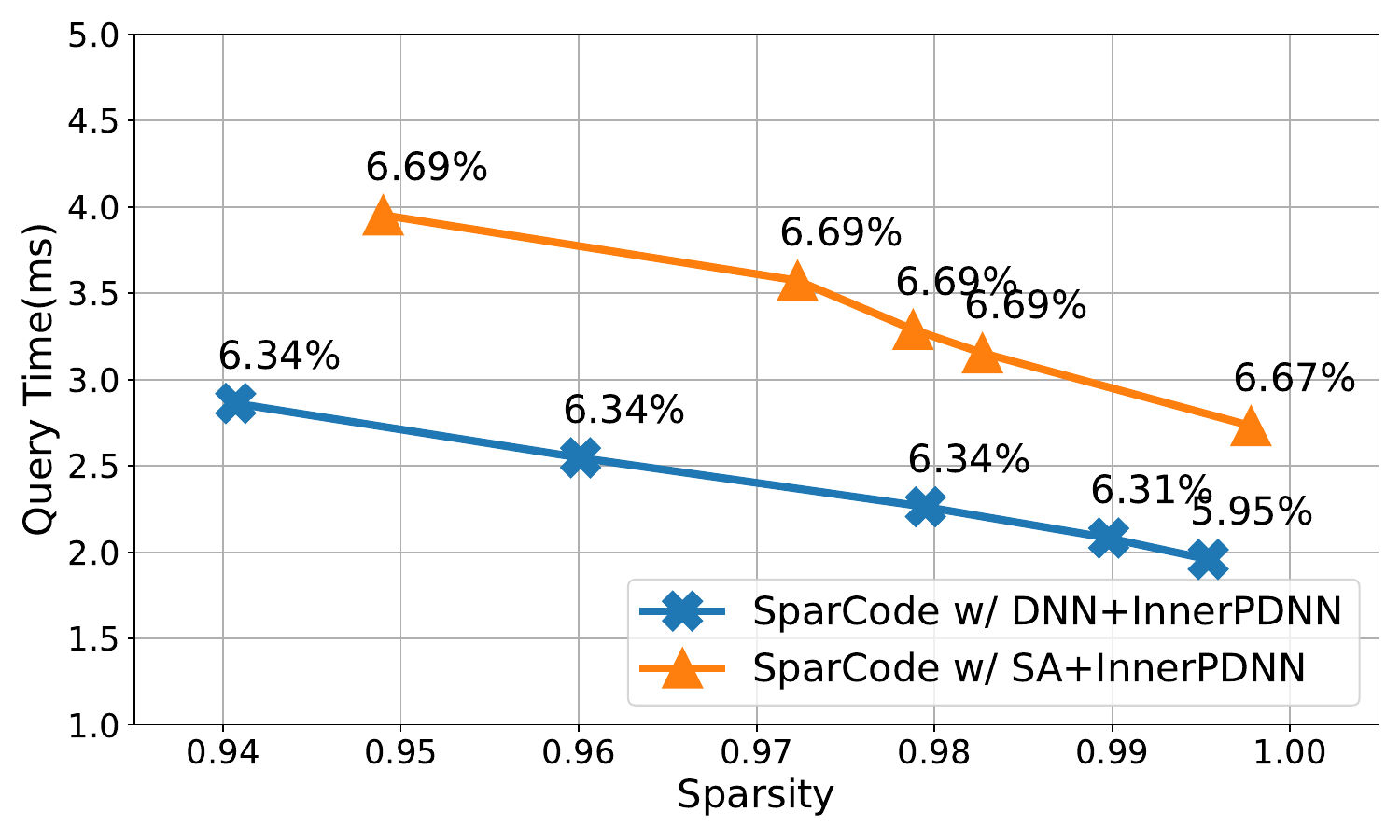}}
	\caption{Effects of sparsity on performance and query time.}
	\label{sparsity_and_performance_and_time}
\end{figure*} 

\textbf{Effect of Sparsity on Performance}. We control sparsity by adjusting the bias $\widetilde{\textbf{b}}$ in Eq.~\ref{sparse_control}. The experimental results of the association of sparsity and performance are shown in Figure~\ref{sparsity_and_performance_and_time}(a) and Figure~\ref{sparsity_and_performance_and_time}(b), which is surprising. For Deezer, given $M$=2 and $N$=256, the setting with 99\% sparsity achieves comparable performance with 0\% sparsity on Recall and NDCG. This result also validates our observation that each code is highly correlated with only a small number of candidates. For memory-limited scenarios, the compressed model with a 99.947\% sparsity can still exceed the performance of two-tower models (refer to Table~\ref{performance_deezer}). For Movielens-10M, such conclusion still holds, with only 1\% scores cached to support a significant accuracy improvement. The above experimental results indicate SparCode is quite capable of online deployment even under extreme conditions.

\textbf{Effect of Sparsity on Inference Speed}. With performance guaranteed, another concern is the inference speed. Here, we are mainly considering the time taken to process each retrieval request, excluding pre-calculation or pre-loading time. We simulate processing a real request in real scenarios. Given the resource constraints, query tower inference runs on the GPU, while the process of retrieving a top-\textit{k} ANN runs on the CPU. In order to respond requests quickly, only one or a small number of requests are processed per inference. We set the batch size as 1 for comparison. 

Figure~\ref{sparsity_and_performance_and_time}(c) compares the inference speed between SparCode and TTMs. 
As shown in Figure~\ref{sparsity_and_performance_and_time}(c), SparCode has similar query time with TTMs with FlatIP index, but yields significantly better performance. For two tower models,  we utilize PQ methods such as IVFPQ index to get lower query time, with slight performance loss for comparison. It is worth noting that SparCode is implemented entirely in python under our experiment settings. While there is a bunch of open-source search engines which supports inverted index searching with much faster speed, SparCode can utilize these engines to achieve better query time, similar to PQ-enhanced TTM, with little performance loss. 

Figure~\ref{sparsity_and_performance_and_time}(d) shows the effects of different sparsity on query time. With the sparsity increases, the query time decreases respectively. The evaluation performance also decreases as we have discussed. 

\begin{table}[t!]
\caption{Performance comparison of SparCode and TDM}
\resizebox{0.99\linewidth}{!}{
\begin{tabular}{c|c|cccc}
\hline
Model    & Interaction Type & R@5    & R@10   & R@20   & R@50   \\ \hline
TDM-DNN  & Dot Product      & 0.40\% & 0.79\% & 1.53\% & 3.71\% \\
TDM-DNN  & MLPs             & 0.64\% & 1.21\% & 2.31\% & 5.09\% \\ \hline
SparCode-DNN & MLPs             & \textbf{0.94\%} & \textbf{1.75\%} & \textbf{3.12\%} & \textbf{6.55\%} \\ \hline
\end{tabular}
}
\label{tab:tdm}
\end{table}

\subsection{Comparsion with TDM.}

TDM~\cite{TDM} is a representative tree-based matching model which, like SparCode, supports advanced interaction between user and item. In this section, we compare SparCode and TDM in terms of model design and performance. We trained two YoutubeDNN models with interactions of dot product and MLPs on Deezer (TT-SVD), respectively, and trained and updated the index of the tree structure based on the trained YoutubeDNN models following the guidance of TDM. The results of SparCode and TDM experiments are shown in the Table~\ref{tab:tdm}, where SparCode-DNN denotes SparCode with YoutubeDNN with the same parameters as TDM.

According to Table~\ref{tab:tdm}, we have the following observations: (a) The TDM-DNN with MLPs is better than that with Dot Product, where R@20 and R@50 improved from 1.53\% to 2.31\% and from 3.71\% to 5.09\%, respectively. This improvement mainly comes from the interaction method, which again validates the limited feature interaction capability of dot product. (b) Compared to TDM-DNN with MLPs, SparCode-DNN has an average relative improvement of 35.07\% in Recall metric, which reflects SparCode's accuracy advantage in model architecture.

Despite their ability to support advanced interaction methods, SparCode and TDM have many differences. First, SparCode employs a sparse inverted index with $O(1)$ inference complexity, while TDM uses a tree structure index with $O(Log(N))$ inference complexity. That is, SparCode only needs to perform a top-\textit{k} retrieval once, while TDM needs to perform $O(Log(N))$ times. Secondly, SparCode is trained end-to-end, while TDM performs both model and index training. In addition, SparCode is easily controllable for the size of inverted index structures, whereas TDM usually needs to keep full binary tree indexes.





%% file: sections/5_conclusion.tex
\section{Conclusion}
 In this paper, we summarize two limitations of the two-tower model: limited feature interaction capability and reduced accuracy in online serving. To address the two limitations, we proposed a new matching paradigm SparCode for improving both recommendation accuracy and inference efficiency. By linking vector quantization and sparse inverted indexing, SparCode introduces an all-to-all interaction module to achieve fine-grained interaction between user and item features and is able to maintain efficient retrieval with $O(1)$ complexity. In addition, we further design the sparse fraction function to control the size of the index structure, so as to reduce the storage pressure. Extensive experimental results on two public datasets show that SparCode has far superior performance and comparable efficiency to the two-tower matching. In the future, we will further explore the application of SparCode to other tasks.

\begin{acks}
This work was supported in part by the National Natural Science Foundation of China (61972219,62276154), the Research and Development Program of Shenzhen (JCYJ20190813174403598, \\ JCYJ20210324120012033), the Overseas Research Cooperation
Fund of Tsinghua Shenzhen International Graduate School (HW2021013), Aminer ShenZhen Scientific Super Brain. We gratefully acknowledge the support of Mindspore\footnote{\url{https://www.mindspore.cn}}, which is a new deep learning computing framework.
\end{acks}